\documentclass[11pt,epsf]{article}

\usepackage{epsfig}
\usepackage{amssymb}
\usepackage{graphicx}
\usepackage{color}
\usepackage{subfigure}

\begin{document}

\baselineskip 6mm
\renewcommand{\thefootnote}{\fnsymbol{footnote}}


\newcommand{\nc}{\newcommand}
\newcommand{\rnc}{\renewcommand}



\newcommand{\tcb}{\textcolor{blue}}
\newcommand{\tcr}{\textcolor{red}}
\newcommand{\tcg}{\textcolor{green}}


\def\be{\begin{equation}}
\def\ee{\end{equation}}
\def\ba{\begin{array}}
\def\ea{\end{array}}
\def\bea{\begin{eqnarray}}
\def\eea{\end{eqnarray}}
\def\nn{\nonumber\\}


\def\ct{\cite}
\def\la{\label}
\def\eq#1{Eq. (\ref{#1})}


\def\a{\alpha}
\def\b{\beta}
\def\g{\gamma}
\def\G{\Gamma}
\def\d{\delta}
\def\D{\Delta}
\def\ep{\epsilon}
\def\e{\eta}
\def\ph{\phi}
\def\Ph{\Phi}
\def\ps{\psi}
\def\Ps{\Psi}
\def\k{\kappa}
\def\l{\lambda}
\def\L{\Lambda}
\def\m{\mu}
\def\n{\nu}
\def\th{\theta}
\def\Th{\Theta}
\def\r{\rho}
\def\s{\sigma}
\def\S{\Sigma}
\def\ta{\tau}
\def\o{\omega}
\def\O{\Omega}
\def\pr{\prime}


\def\half{\frac{1}{2}}

\def\goto{\rightarrow}

\def\na{\nabla}
\def\grad{\nabla}
\def\curl{\nabla\times}
\def\div{\nabla\cdot}
\def\pa{\partial}

\def\bra{\left\langle}
\def\ket{\right\rangle}
\def\lb{\left[}
\def\lc{\left\{}
\def\ls{\left(}
\def\lp{\left.}
\def\rp{\right.}
\def\rb{\right]}
\def\rc{\right\}}
\def\rs{\right)}

\def\vac#1{\mid #1 \rangle}


\def\td#1{\tilde{#1}}
\def\check{ \maltese {\bf Check!}}


\def\Tr{{\rm Tr}\,}
\def\det{{\rm det}}


\def\bc#1{\nnindent {\bf $\bullet$ #1} \\ }
\def\ch {$<Check!>$ }
\def\ss {\vspace{1.5cm}}

\begin{titlepage}

\hfill\parbox{5cm} { }

\vspace{25mm}

\begin{center}
{\Large \bf Conductivity in the anisotropic background}

\vskip 1. cm
  { Bum-Hoon Lee$^{a,b}$\footnote{e-mail : bhl@sogang.ac.kr},
  Siyoung Nam$^a$\footnote{e-mail : ¢Óstringphy@gmail.com},
  Da-Wei Pang$^b$\footnote{e-mail : pangdw@sogang.ac.kr} and
  Chanyong Park$^b$\footnote{e-mail : cyong21@sogang.ac.kr}}

\vskip 0.5cm

{\it $^a\,$Department of Physics, Sogang University, Seoul 121-742, Korea}\\
{ \it $^b\,$Center for Quantum Spacetime (CQUeST), Sogang University, Seoul 121-742, Korea }\\

\end{center}

\thispagestyle{empty}

\vskip2cm


\centerline{\bf ABSTRACT} \vskip 4mm

\vspace{1cm}
By using the gauge/gravity duality, we investigate the dual field theories of the
anisotropic backgrounds,
which are exact solutions of Einstein-Maxwell-dilaton theory with a
Liouville potential. When we turn on the bulk gauge field fluctuation $A_{x}$
with a non-trivial dilaton coupling,
the AC conductivity of this dual field theory is proportional to the frequency
with an exponent depending on parameters of the anisotropic background.
In some parameter regions, we find that this conductivity can have the negative exponent
like the strange metal. In addition, we also investigate another $U(1)$ gauge field fluctuation,
which is not coupled with a dilaton field. We classify all possible
conductivities of this system and find that the exponent of
the conductivity is always positive.

\vspace{2cm}


\end{titlepage}

\renewcommand{\thefootnote}{\arabic{footnote}}
\setcounter{footnote}{0}

\tableofcontents


\section{Introduction}
The AdS/CFT correspondence\cite{Maldacena:1997re, Aharony:1999ti},
which relates the dynamics of strongly coupled field theories to the
corresponding dual gravity theories, has provided us a powerful tool
for studying physical systems in the real world. Recently, inspired
by condensed matter physics, the applications of the AdS/CFT
correspondence to condensed matter physics(sometimes called the
AdS/CMT correspondence) have been accelerated enormously. As a
strong-weak duality, the AdS/CFT correspondence makes it possible to
investigate the strongly coupled condensed matter systems in the
dual gravity side. Therefore it is expected that we can acquire
better understandings for certain condensed matter systems via the
AdS/CFT correspondence. Some excellent reviews are given
by~\cite{Hartnoll:2009sz}.

According to the AdS/CFT correspondence, once the dual boundary
field theory lies at finite temperature, there should exist a corresponding black
hole solution in the bulk. Then some properties, like various conductivities
\cite{Gubser:2008wz,Hartnoll:2007ai}, superconductor \cite{Horowitz:2009ij,Hartnoll:2008kx}
and non-fermi liquid \cite{Lee:2008xf,Liu:2009dm}, of the dual field
theory can be inferred from the black hole. One particular class of
such black hole solutions is a charged dilaton black
hole \cite{Gibbons:1987ps, Preskill:1991tb, Garfinkle:1990qj,
Holzhey:1991bx}, where the gauge coupling of the Maxwell term
is governed by a dilaton field $\phi$. Charged dilaton
black holes in the presence of a Liouville potential were studied
in Ref. \cite{Cai:1996eg}. Such charged dialton black hole solutions
possess two interesting properties. First, for certain specific
values of the gauge coupling, the charged dilaton black holes can be
embedded into string theory. Second, the entropy vanishes in the
extremal limit, which may signify that the thermodynamic description
breaks down at extremality. Such peculiar features suggest that
their AdS generalizations may provide interesting holographic
descriptions of condensed matter systems.

Recently, holography of charged dilaton black holes in AdS$_4$ with
planar symmetry was extensively investigated
in~\cite{Goldstein:2009cv}. The near horizon geometry was
Lifshitz-like with a dynamical exponent $z$ determined by the
dilaton coupling. The global solution was constructed via numerical
methods and the attractor behavior was also discussed. The authors
also examined the thermodynamics of near extremal black holes and
computed the AC conductivity in zero-temperature background. For
related works on charged dilaton black holes see~\cite{Gubser:2009qt,
Gauntlett:2009bh, Cadoni:2009xm, Chen:2010kn}.

In this paper we focus on conductivities of charged dilaton black
hole solutions \cite{Taylor:2008tg} with a Liouville potential at both zero and
finite temperature. At first we obtain exact solutions, both extremal
and non-extremal, of the Einstein-Maxwell-dilaton theory, where the
scalar potential takes the Liouville form. The extremal solution
possesses anisotropic scaling symmetry which reduces to the Lifshitz-like
metric \cite{Kachru:2008yh} in a certain limit.
In Ref. \ct{Goldstein:2009cv,Charmousis:2010zz}, after inserting an irrelevant operator
which deforms the asymptotic geometry to $AdS_{4}$, the electric conductivity
was calculated. In this paper, we concentrate on the undeformed geometry
and find exact solution describing undeformed geometry. Secondly we calculate the electric
conductivity by considering the fluctuation of the gauge field
$A_{x}$. The corresponding equation of motion for $A_{x}$ can be
transformed into a Schr\"{o}dinger equation, which enables us to
evaluate the conductivity easily~\cite{Horowitz:2009ij}.
In the usual (deformed) AdS spaces, because there is no non-trivial dilaton coupling to the gauge field
the conductivity of the dual theory is usually proportional to
the frequency with the non-negative exponent, which is opposite to the strange metallic one.
In the anisotropic background
the electric conductivity generally is proportional to the frequency with an exponent determined by
the non-trivial dilaton coupling and other background geometric parameters. Furthermore, in
appropriate parameter regions we can find the strange metal-like conductivity proportional
to the frequency with the negative exponent.

Next we introduce another $U(1)$ gauge field, which does not have a coupling
with the dilaton field, and calculate the conductivity by considering
the fluctuations of this new gauge field. There are some motivations for considering
new gauge field with different dilaton coupling. First, if our four-dimensional
gravity theory is originated from the ten dimensional string theory,
there exist other many gauge fields having different dilaton coupling depending
on how the string theory is compactified. Second, if we promote this bottom-up
approach to the top-down approach, the gauge field living on the probe brane world
volume typically has a dilaton coupling like $e^{- \ph}$.
So studying the dynamics of the bulk gauge fluctuations having various
dilaton couplings would shed light on understanding dual gauge theories.
In this paper, we concentrate on a special case in which the gauge field fluctuation has
no dilaton coupling and calculate the electric conductivities of the dual system.
We find that the exponent of the conductivity for all
parameter regions is always positive.
This is the main difference between the dual theory with or without a
non-trivial dilaton coupling, which implies that
that the non-trivial dilaton coupling
is very important to explain the strange metallic behavior.

The rest of the paper is organized as follows: in Section 2 we find
the exact solutions of the charged dilaton black holes with a
Liouville potential, both extremal and non-extremal. We calculate
the conductivity by turning on the gauge field fluctuations coupled with the dilaton field in
Section 3. In Section 4
we calculate the conductivity induced by a new $U(1)$ gauge field, which is not coupled with
the dilaton field.
Summary and discussion will be given in the final section. \\

{\bf Note added}: In the final stage of this work, we noticed that
similar solutions were studied extensively~\cite{Charmousis:2010zz}.
However, one key difference is that they required the solutions to
be asymptotically $AdS$ while we treat the solutions to be global.

\section{Anisotropic background with a Liouville potential}

In the real condensed matter theory systems containing anisotropic scaling commonly appear.
To describe this system holographically the Lifshitz background has been
studied by many authors \cite{Kachru:2008yh} by using the generalized AdS/CFT correspondence,
so called the gauge/gravity duality.
Here, we consider a different background having more general anisotropic
scaling and investigate physical quantities of the dual field theory.

We start with the following action
\begin{equation}    \la{orgact}
S=\int
d^{4}x\sqrt{-g}[R-2(\nabla\phi)^2-e^{2\alpha\phi}F_{\mu\nu}F^{\mu\nu}-V(\ph)],
\end{equation}
where $\ph$ and $V(\ph)$ represent a dilaton field and its potential.
Notice that the non-trivial dilaton coupling in front of the Maxwell term
can provide a different physics from the ordinary relativistic
field theory dual to AdS space. So main goal of the present work is to investigate the dilaton coupling
effect on the physical quantity like the conductivity of the dual field theory.
Equations of motion for metric $g_{\m\n}$, dilaton field and U(1) gauge field are
\begin{eqnarray}
R_{\mu\nu}-\frac{1}{2}Rg_{\mu\nu}+\frac{1}{2}g_{\mu\nu}V(\phi) &=& 2\partial_{\mu}\phi
\partial_{\nu}\phi-g_{\mu\nu}(\nabla\phi)^{2}+2e^{2\alpha\phi}F_{\mu\lambda}{F_{\nu}}^{\lambda}
-\frac{1}{2}g_{\mu\nu}e^{2\alpha\phi}F^{2}, \\
\partial_{\mu}(\sqrt{-g}\partial^{\mu}\phi) &=& \frac{1}{4}\sqrt{-g}\frac{\partial
V(\phi)}{\partial\phi}+\frac{\alpha}{2}\sqrt{-g}e^{2\alpha\phi}F^{2},~~~ \\
0 &=& \partial_{\mu}(\sqrt{-g}e^{2\alpha\phi}F^{\mu\nu}) \la{eqgauge}.
\end{eqnarray}
Now, we choose a Liouville-type potential as a dilaton potential \be
V(\phi)=2\Lambda e^{- \eta\phi} . \ee For $\eta=0$, the dilaton
potential reduces to a cosmological constant, which was studied in
Ref. \cite{Goldstein:2009cv}. To solve equations of motion, we use
the following ansatz corresponding to a zero temperature solution
\begin{equation}
ds^{2}=-a(r)^{2}dt^{2}+\frac{dr^{2}}{a(r)^{2}}+b(r)^{2}(dx^{2}+dy^{2}) ,
\end{equation}
with
\begin{equation}
a(r)=t_{0}r^{a_{1}},~~~b(r)=b_{0}r^{b_{1}},~~~\phi(r)=-k_{0}\log r.
\end{equation}
If we turn on a time-component of the gauge field $A_t$ only, from the above metric ansatz
the electric flux satisfying \eq{eqgauge} becomes
\begin{equation}
F_{tr}=\frac{q}{b(r)^{2}}e^{-2\alpha\phi}.
\end{equation}
The rest of equations of motion are satisfied when various
parameters appearing in the above are given by
\bea    \la{param}
 &&
a_{1}=1+\frac{k_{0}}{2}\eta,~~~b_{1}=\frac{(2\alpha-\eta)^{2}}{(2\alpha-\eta)^{2}+16},
~~~k_{0}=\frac{4(2\alpha-\eta)}{(2\alpha-\eta)^{2}+16}, ~~~b_{0}=1 ,
\nn &&
t_{0}^{2}=\frac{-2\Lambda}{(a_{1}+b_{1})(2a_{1}+2b_{1}-1)},~~~
q^{2}=- \ls \frac{2k_{0}}{a_{1}+b_{1}}+\frac{\eta}{2} \rs
\frac{\L}{\a} ,
\eea
where $\L$ is negative.
This solution described by three free parameters $\a$, $\eta$ and $\L$
is an exact solution of equations of motion. Notice that $b_1$ in \eq{param} is
always smaller than $1$. For $\eta=0$ and $\L = -3$ this
solution reduces to the one in Ref. \cite{Goldstein:2009cv}, as
previously mentioned. For $2\a = \eta$, the above solution becomes
$AdS_2 \times R^2$. If we take a limit $\a \to \infty$ and at the
same time set $q=\eta=0$, we can obtain $AdS_4$ geometry. When
$\eta$ is proportional to $\a$ like $\eta = c \a$, the metric in the
limit, $\a \to \infty$, reduces to a Lifshitz-like one
\be
ds^2 = - t_0^2 r^{2 z}
dt^2 + \frac{dr^2}{t_0^2 r^{2z}} + r^2 \ls dx^2 + dy^2  \rs , \ee
with \be z = \frac{2+c}{2-c} .
\ee
The exponent $z$ is given by $2$ for
$c=2/3$ and $3$ for $c=1$, etc.

The above zero temperature solution can be easily extended to a finite temperature one
describing a black hole.
With the same parameters in \eq{param}, the black hole solution becomes
\be \la{anisomet}
ds^{2}=-a(r)^{2}f(r)dt^{2}+\frac{dr^{2}}{a(r)^{2}f(r)}+b(r)^{2}(dx^{2}+dy^{2}),
\ee
where
\be
f(r)=1-\frac{r^{2a_{1}+2b_{1}-1}_{h}}{r^{2a_{1}+2b_{1}-1}} .
\end{equation}
Notice that since the above black hole factor does not include U(1) charge
this solution corresponds to not a Reissner-Nordstr\"{o}m but Schwarzschild black hole.
The Hawking temperature of this black hole is given by
\be
T \equiv  \lp \frac{1}{4 \pi} \frac{\pa ( a(r)^2 f)}{\pa r} \right|_{z=z_h}
= \frac{(2 a_1 + 2 b_1 -1) t_0^2  r_h^{2 a_1 -1} }{4 \pi} ,
\ee
where $r_h$ means the position of the black hole horizon.

\section{Fluctuation of background gauge field $A_x$}

In this section, we consider the gauge field fluctuation $A_x$ and calculate the electric
conductivity of the dual theory. For convenience, we introduce new coordinate variable
$u=r^{b_1}$. Then, the metric in \eq{anisomet} becomes
\be
ds^2 = - g(u) f(u) e^{- \chi(u)} dt^2 + \frac{du^2}{g(u) f(u)} + u^2 \ls dx^2  + dy^2 \rs ,
\ee
where
\bea
g(u) &=& t_0^2 b_1^2 u^{2 (a_1 + b_1 -1)/b_1} , \nn
e^{\chi(u)} &=& b_1^2 u^{2(b_1 -1)/b_1} ,
\eea
with the black hole factor
\be
f(u) = 1-\frac{u_h^{(2 a_1 + 2 b_1 -1 )/b_1} }{u^{(2 a_1 +2 b_1 -1)/b_1}} .
\ee

On this background, we turn on the gauge field fluctuation $A_x$ together with the
metric fluctuation $g_{tx}$. Although there is another relevant metric fluctuation $g_{ux}$, in the
$g_{ux} = 0$ and $A_u=0$ gauge
only $A_x$ and $g_{tx}$ describe the vector fluctuations of the bulk theory.
From the action for the gauge fluctuation
\be
\d S = - \frac{1}{4} \int d^4 x \sqrt{- g} \ h^2  F_{\m\n}  F^{\m\n} ,
\ee
where $h^2 = 4 e^{2 \a \ph}$ is a coupling function, the equation of motion for $A_x$
becomes
\be
0 = \frac{1}{\sqrt{-g}} \pa_{\m} \ls \sqrt{-g} \ h^{2}
g^{\m \r} g^{x \n} F_{\r \n} \rs .
\ee
Under the following ansatz,
\bea
A_x (t,u) &=& \int \frac{d w}{ 2 \pi} e^{- i w t } A_x (w,u) , \nn
g_{tx} (t,u) &=& \int \frac{d w}{ 2 \pi} e^{- i w t }  g_{tx} (w,u) ,
\eea
the above equation for $A_x$ can be rewritten as
\be \la{eqforbg}
0 = \pa_u \ls e^{- \frac{\chi}{2}} g f h^2 \pa_u A_x \rs
+ w^2 e^{\frac{\chi}{2}} \frac{h^2}{g f} A_x
+ e^{\frac{\chi}{2}} h^2 (\pa_u A_t) \ls g_{tx}' - \frac{2}{u} g_{tx}\rs
\ee
and the $(u,x)$-component of Einstein equation is given by
\be \la{eqformet}
g_{tx}' - \frac{2}{u} g_{tx} = - h^2 (\pa_u A_t) A_x .
\ee
Combining above two equations gives
\be \la{eqfl}
0 = \pa_u \ls  e^{- \frac{\chi}{2}} g f h^2 \pa_u A_x \rs
+ w^2 e^{\frac{\chi}{2}} \frac{h^2}{g f} A_x
- e^{\chi/2} h^4 (\pa_u A_t)^2 A_x .
\ee

Introducing a new variable and new wave function
\bea    \la{nnvar}
- \frac{\pa}{\pa v} &=& e^{- \frac{\chi}{2}} g \frac{\pa}{\pa u} , \nn
A_x &=& \frac{\Ps}{ \sqrt{f} h},
\eea
\eq{eqfl} simply reduces to a Schr\"{o}dinger-type equation
\be \la{sheq}
0 = \Ps'' + V(v) \Ps
\ee
with the effective potential
\be \la{effpot}
V(v) = \ls w^2 + \frac{(f')^2}{4} \rs \frac{1}{f^2}
- \ls \frac{f' h'}{h} + \frac{f''}{2} + \frac{h^2}{g} e^{\chi} (A_t')^2  \rs \frac{1}{f}
- \frac{h''}{h} ,
\ee
where the prime implies a derivative with respect to $v$. Using the first equation
in \eq{nnvar} we can easily find $u$ as function of $v$.
Here, we will concentrate on the case $2 a_1 > 1$, in which
$v$ is given by
\be \la{chcoord}
v = \frac{1}{(2 a_1 -1) \ t_0^2 \ u^{\frac{2 a_1 -1}{b_1}}} .
\ee
In $v$-coordinate the boundary ($u=\infty$) is located at $v=0$.
In the black hole geometry, the black hole horizon $v_h$ is given by
\be
v_h = \frac{1}{(2 a_1 -1) \ t_0^2 \ u_h^{\frac{2 a_1 -1}{b_1}}} .
\ee
So the zero temperature corresponds to put the black hole horizon to $v=\infty$ ($u=0$).

\subsection{At zero temperature}

We first consider the zero temperature case, in which $f=1$ and $f'=f''=0$.
The effective potential \eq{effpot} reduces to
\be
V(v) = w^2 -  \frac{h^2}{g} e^{\chi} (A_t')^2 - \frac{h''}{h} .
\ee
Using \eq{nnvar}, the above effective potential
can be written as
\be \la{efp}
V(v) = w^2 - \frac{c}{v^2} ,
\ee
with a constant $c$
\be \la{wtgtx}
c = \frac{4 (16 + 4 \a^2 - \eta^2) \lb 8 + (2 \a - \eta) (\a
+ \eta)\rb }{ (16 + 4 \a^2 + 4 \a \eta - 3 \eta^2)^2},
\ee
where we use \eq{param}.
The exact solution $\Ps$ of the Schr\"{o}dinger equation becomes
\be \la{soA0}
\Ps = c_1 \sqrt{v} H^{(1)}_{\d} (w v) + c_2 \sqrt{v} H^{(2)}_{\d}(w v) ,
\ee
where $H^{(i)}$ implies the $i$-th kind of Hankel function and
\be \la{pprel}
\d = \frac{\sqrt{1+ 4 c}}{2} .
\ee
At the horizon ($v=\infty$),
the first or second term in \eq{soA0} satisfies the incoming or outgoing boundary
condition respectively. So to pick up the solution satisfying the incoming boundary
condition we should set $c_2 = 0$. Then, the solution has the following expansion near the boundary
\be
\Ps \approx \Ps_0 \ls  v^{\half - \d}
-  \ls \frac{w}{2} \rs^{2 \d} \frac{\G (1 - \d)}{\G (1+\d)} e^{- i \pi \d} v^{\half+ \d} \rs
\ee
with
\be
c_1 = \frac{i \pi }{\G (\d)} \ls \frac{w}{2} \rs^{\d} \Ps_0 .
\ee
From this result together with \eq{nnvar}, $A_x$ at the boundary ($u=\infty$) becomes
\be
A_x = A_0 \ls  u^{\lb \a k_0 - (2 a_1 -1)(\half - \d) \rb /b_1} -
\td{c} \ u^{\lb \a k_0 - (2 a_1 -1)(\half + \d) \rb /b_1} \rs ,
\ee
where
\bea
A_0 &=& \frac{\Ps_0}{2 \lb (2 a_1 -1) t_0^2 \rb^{\half-\d}} , \nn
\td{c} &=& \ls \frac{w}{2} \rs^{2 \d} \frac{\G (1 - \d)}{\G (1+\d)}
\frac{e^{- i \pi \d}}{\lb (2 a_1 -1) t_0^2 \rb^{2 \d}} .
\eea
The boundary action for the gauge fluctuation $A_x$ becomes
\bea
S_B &=& - \half \int d^3 x \ \sqrt{-g} \ h^2 g^{uu} g^{xx} A_x \ \pa_u A_x \nn
&=& \half \int d^3 k \ 4 t_0^2 \td{c} \lb (2 a_1 -1)(\half + \d) - \a k_0 \rb A_0^2 ,
\eea
so that the Green function is given by
\be
G_{xx} \equiv \frac{\pa^2 S_B}{\pa A_0 \pa A_0}
= 4 t_0^2  \lb (2 a_1 -1)(\half + \d) - \a k_0  \rb \td{c}  \sim  w^{2 \d} .
\ee
Finally, the AC conductivity of the dual system reads off
\be \la{guide0}
\s = \frac{G_{xx}}{i w} \sim  w^{2 \d - 1} .
\ee
For the DC conductivity we should set $w = 0$. Then, the DC conductivity
of this system becomes infinity for $2 \d < 1$ or zero for $2 \d > 1$.
In the real world, there exist some condensed matter systems having the AC conductivity
with a negative exponent like the strange metal \cite{vandeMarel:2003wn,Hartnoll:2009ns}.
If we choose $2 \d -1 = - 0.65$.
the AC conductivity of this system can describes the strange metallic behavior
$\s \sim w^{-0.65}$. Notice that since there are two free parameters, $\a$ and $\eta$, for $\L=-3$
the strange metallic behavior can appear in infinitely many parameter regions.
For example, the following parameters,
$(\a,\eta) \approx (1,3.804), \ (2,5.196), \ (2,5.338), \cdots$,
satisfy $\s \sim w^{-0.65}$
and give the regular value, $q^2 > 0$.
As will be shown in Sec. 4, if we consider the gauge field fluctuation without
the dilaton coupling the conductivity becomes a constant when the spatial momentum is zero.
If turning on the spatial momentum, the conductivity always grows up
as the frequency increases.
These behaviors are different with those of the strange metal, which implies that
the dilaton coupling of the gauge field plays an important role
for describing the strange
metallic behavior holographically. In the next section, we will show the strange metallic behavior
at finite temperature.

\subsection{At finite temperature}

We consider the gauge and metric fluctuations
at the finite temperature background.
At the horizon, the dominant term in the effective potential in \eq{effpot} is
\be
U(v) \approx \ls w^2 + \frac{(f_h')^2}{4} \rs \frac{1}{f_h^2}
= \ls w^2 +   \frac{(2 a_1 + 2 b_1 - 1)^2}{4 (2 a_1 -1)^2 v_h^2} \rs \frac{1}{f_h^2} ,
\ee
where $f_h$ means the value of $f$ at the horizon. Then, the approximate solution
is given by
\be \la{softd}
\Ps = c_1 f^{\n_-} + c_2 f^{\n_+} ,
\ee
where
\be
\n_{\pm} = \half \pm i \sqrt{ w^2 + \frac{(2 a_1 + 2 b_1 - 1)^2}{4 (2 a_1 -1)^2 v_h^2}
- \frac{1}{4} } .
\ee
Notice that at the horizon the first and second term in \eq{softd} satisfy the
incoming and outgoing boundary condition respectively. Imposing the incoming boundary
condition we can set $c_2 = 0$.

\begin{figure}
\begin{center}
\vspace{-1cm}
\hspace{-0.5cm}
\subfigure{ \includegraphics[angle=0,width=0.45\textwidth]{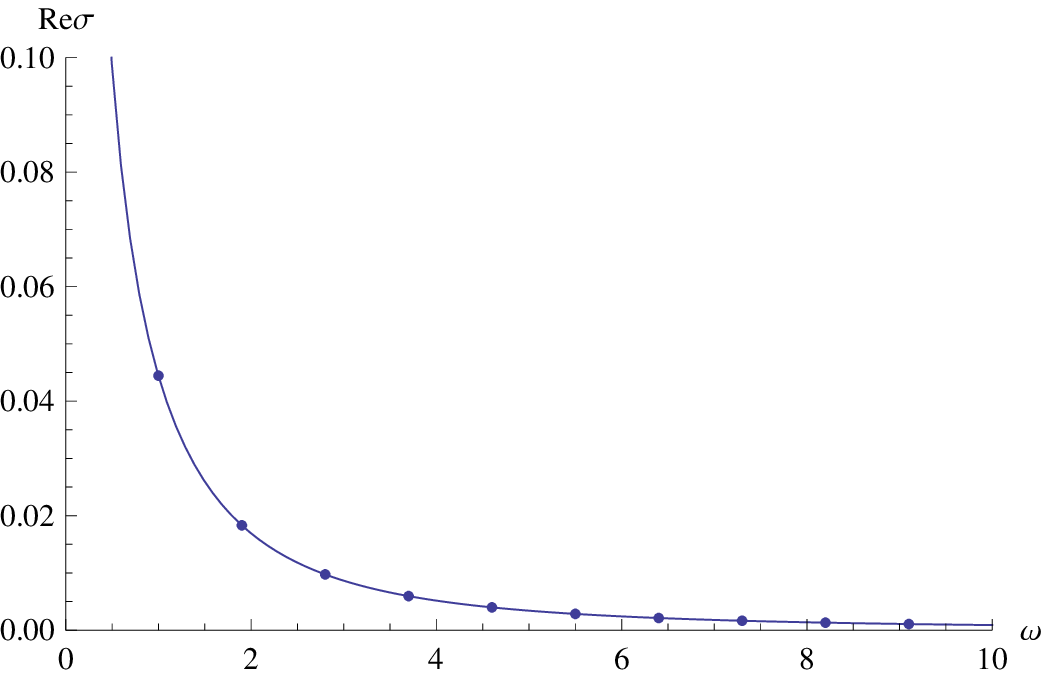}}
\hspace{-0cm}
\subfigure{ \includegraphics[angle=0,width=0.45\textwidth]{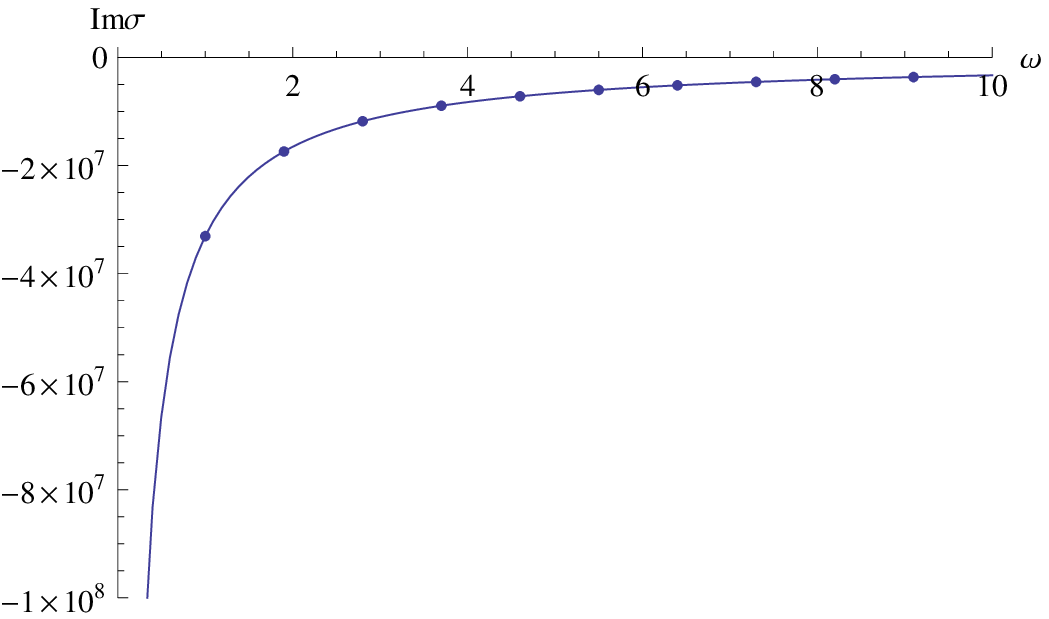}}
\vspace{-0cm} \\
\caption{\small The conductivity at the finite temperature where we choose
$\a=2$, $\eta=1$ and $\L=-3$.}
\label{ConAtFWC}
\end{center}
\end{figure}

Now, we investigate the asymptotic behavior of $\Ps$. Since the leading behavior of
the effective potential near the boundary is given by \eq{efp}, the perturbative
solution can be described by
\be
\Ps = d_1 \ v^{\half - \d} + d_2 \ v^{\half + \d},
\ee
where $\d$ has been defined in \eq{pprel} together with \eq{wtgtx}.
By solving \eq{sheq} numerically, we can determine the numerical value of $d_1$ and
$d_2$. In $u$-coordinate, the gauge field fluctuation has the following perturbative
form
\be
A_x = A_0 \ls u^{\lb a k_0 - (2 a_1 -1 ) (\half - \d) \rb /b_1} +
\frac{d_2}{d_1 \lb (2 a_1 - 1) t_0^2 \rb^{2 \d} }
u^{\lb a k_0 - (2 a_1 -1 ) (\half + \d) \rb /b_1} \rs ,
\ee
where to determine the boundary value of $A_x$ as $A_0$ we set
\be
d_1 = 2 \lb (2 a_1 - 1) t_0^2 \rb^{\half - \d} A_0 .
\ee
The boundary action for $A_x$ becomes
\be
S_B = \frac{A_0^2}{2} \int d^3 x \ \frac{4 t_0^2 d_2 \lb
(2 a_1 -1) (\half + \d) + b_1   -  \a k_0 \rb}{d_1 \lb (2 a_1 - 1) t_0^2 \rb^{2 \d}} .
\ee
Then, we can easily find the AC conductivity at finite temperature
\be
\s = \frac{4 t_0^2}{i w} \lb (2 a_1 -1) (\half + \d) + b_1   -  \a k_0 \rb
\frac{ d_2 }{d_1 \lb (2 a_1 - 1) t_0^2 \rb^{2 \d}} ,
\ee
where the last part $\frac{ d_2 }{d_1 \lb (2 a_1 - 1) t_0^2 \rb^{2 \d}}$ can be numerically
calculated by solving the Schr\"{o}dinger equation together with
the initial data at the horizon.
In Figure \ref{ConAtFWC}, we plot the real and imaginary AC conductivity.

We can fit the dual AC conductivity of Figure \ref{ConAtFWC}
with the following expected form
\be \la{eqacs}
\s = a \ w^{-b} .
\ee
If we choose $\a=2$, $\eta=1$ and $\L=-3$
for the simple numerical calculation, the conductivity.
can be fitted by \eq{eqacs} with
$a \approx 0.045$ and $b \approx 1.5$.

\section{New U(1) gauge field fluctuation}

As mentioned in Introduction, it is interesting to investigate the conductivity of
the gauge field having different dilaton coupling. Here, we will concentrate on new gauge field fluctuation
having no dilaton coupling \cite{Sin:2009wi}. Even in this case, due to the parameters in the original
action, we can find several different conductivities depending on the
parameter region.

Before starting the calculation for the Green functions and conductivity in various parameter
regions, we introduce a different coordinate $z = 1/r$ for later convenience. In the
$z$-coordinate, the black hole metric is rewritten as
\be
ds^{2}=- \frac{t_0^2}{z^{2 a_1}} f(z) dt^{2}+\frac{ z^{2 a_1} dz^{2}}{t_0^2 z^4 f(z)}
+ \frac{dx^{2}+dy^{2}}{z^{2 b_1}}
\ee
with
\be
f(z) = 1 - \frac{z^{2 a_1 + 2 b_1 -1}} {z_h^{2 a_1 + 2 b_1 -1}},
\ee
where all parameters are same as ones in \eq{param} and $z_h$ implies the event horizon
of the black hole. In this coordinate, the Hawking temperature becomes
\be
T = \frac{(2 a_1 + 2 b_1 -1) t_0^2 }{4 \pi z_h^{2 a_1 -1}} .
\ee

Now, we introduce another U(1) gauge field fluctuation, which is not coupled with a dilaton field
\be
\d S = - \frac{1}{4} \int d^4 x \sqrt{-g} f_{\m\n} f^{\m\n} ,
\ee
where $f_{\m\n} = \pa_{\m} a_{\n} - \pa_{\n} a_{\m}$
and we absorb a gauge coupling constant to the gauge field.
To obtain equations of motion for new gauge field, we first choose $a_z = 0$ gauge and turn on the
$x$-component of the gauge fluctuation only
\be
a_x (x) = \int \frac{d^3 k}{(2 \pi)^3} e^{- i w t + i \vec{k} \cdot \vec{x}} a_x (k,z) ,
\ee
where a vector $\vec{x}$ corresponds to two-dimensional spatial coordinates.
In the comoving frame, $\vec{k} = (w,0,k)$, the equation of this gauge fluctuation becomes
\be    \la{maxwell}
0 =  a_x'' + \ls \frac{2(1-a_1)}{z} + \frac{f'}{f} \rs a_x'
+ \ls \frac{w^2}{t_0^4 z^{4(1- a_1)} f^2}  -
\frac{k^2}{t_0^2 z^{4-2 a_1-2b_1} f} \rs a_x ,
\ee
where the prime implies a derivative with respect to $z$ and $f$ is a black hole factor.

\subsection{At zero temperature}

In this section, we investigate various Green functions at zero temperature,
which can be obtained by setting $f=1$ and $f'=0$.

\subsubsection{for $a_1 \le 1$}

Notice that $b_1$, as mentioned previously,
is always smaller than $1$ and the case, $b_1 = 1$, can be considered
as the limit $\a \to \infty$.

\vspace{1cm}
\noindent{\bf i) $ \half < a_1 = b_1 \le 1$} \\

At first, we consider a simple case $a_1 = b_1 \le 1$. In this case,
the equation governing the transverse gauge field fluctuation becomes
\bea
0 &=& a_x'' +    \frac{2 \d}{z}  a_x' +  \frac{\g}{z^{4 \d}}   a_x,
\eea
with
\be
\d = 1- a_1  \qquad {\rm and} \qquad \g = \frac{w^2}{t_0^4} - \frac{k^2}{t_0^2} ,
\ee
where $0 \le \d < 1$. The exact solution of the above equation is given by
\be
a_x = c_1  \exp(i \frac{\sqrt{\g} z^{1-2 \d}}{1- 2 \d})
+ c_2 \exp(- i \frac{\sqrt{\g} z^{1-2 \d}}{1- 2 \d}) .
\ee
At the horizon ($z \to \infty$), the first or second term satisfies the
incoming or outgoing boundary condition respectively. Imposing the incoming boundary condition,
the solution reduces to
\be \la{soexs}
a_x = c_1  \exp \ls i \frac{\sqrt{\g} }{1- 2 \d} z^{1-2 \d} \rs .
\ee
For $\d > 1/2$, we can not purturbatively expand this solution near the boundary ($z=0$). In
such case it is unclear how to define the dual operator, so we consider only the case $\d < 1/2$
(or $a_1 > 1/2 $) from now on. In this case, $a_x$ has the following expansion near the boundary
\be
a_x = a_0 \ls 1 + i \frac{\sqrt{\g} }{1- 2 \d} z^{1-2 \d} + \cdots \rs ,
\ee
where $a_0= c_1$ corresponds to the boundary value of $a_x$,
which can be identified with the source term
of the dual gauge operator. According to the gauge/gravity duality,
the on-shell gravity action
can be interpreted as a generating functional for the dual gauge operator.
The on-shell gravity action corresponding to the boundary action is given by
\be \la{bdact}
S_B = \half \int d^3 x \ \sqrt{-g} \ g^{zz} g^{xx} a_x \ \pa_z a_x
= \frac{i a_0^2}{2}  \int d^3 k  t_0^2 \sqrt{\g} .
\ee
Using this on-shell action
we can easily calculate the Green function by varying the on-shell action with respect to the
source.

\begin{figure}
\begin{center}
\vspace{1cm}
\hspace{-0.5cm}
\subfigure{ \includegraphics[angle=0,width=0.5\textwidth]{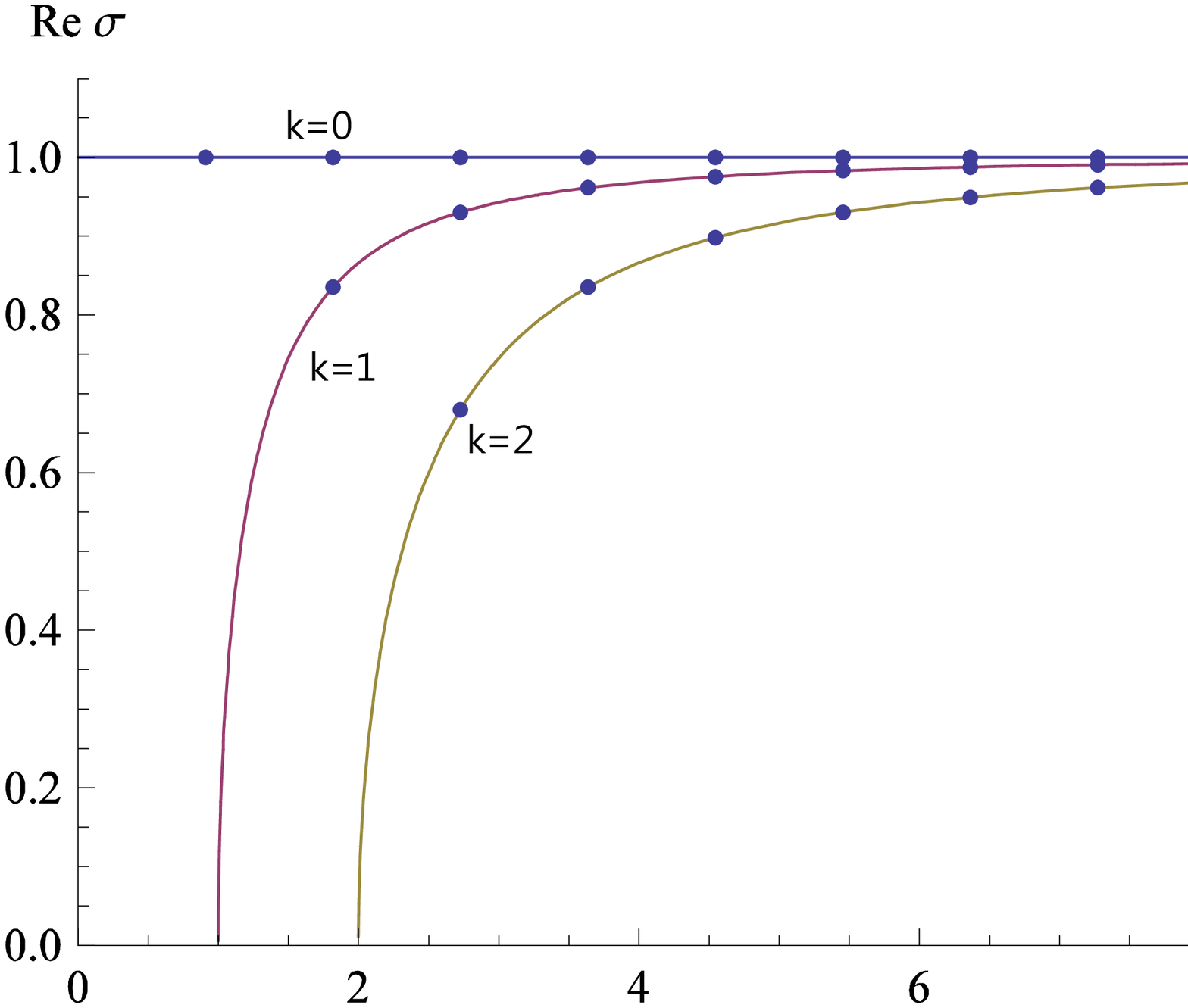}}
\hspace{-0.5cm}
\subfigure{ \includegraphics[angle=0,width=0.5\textwidth]{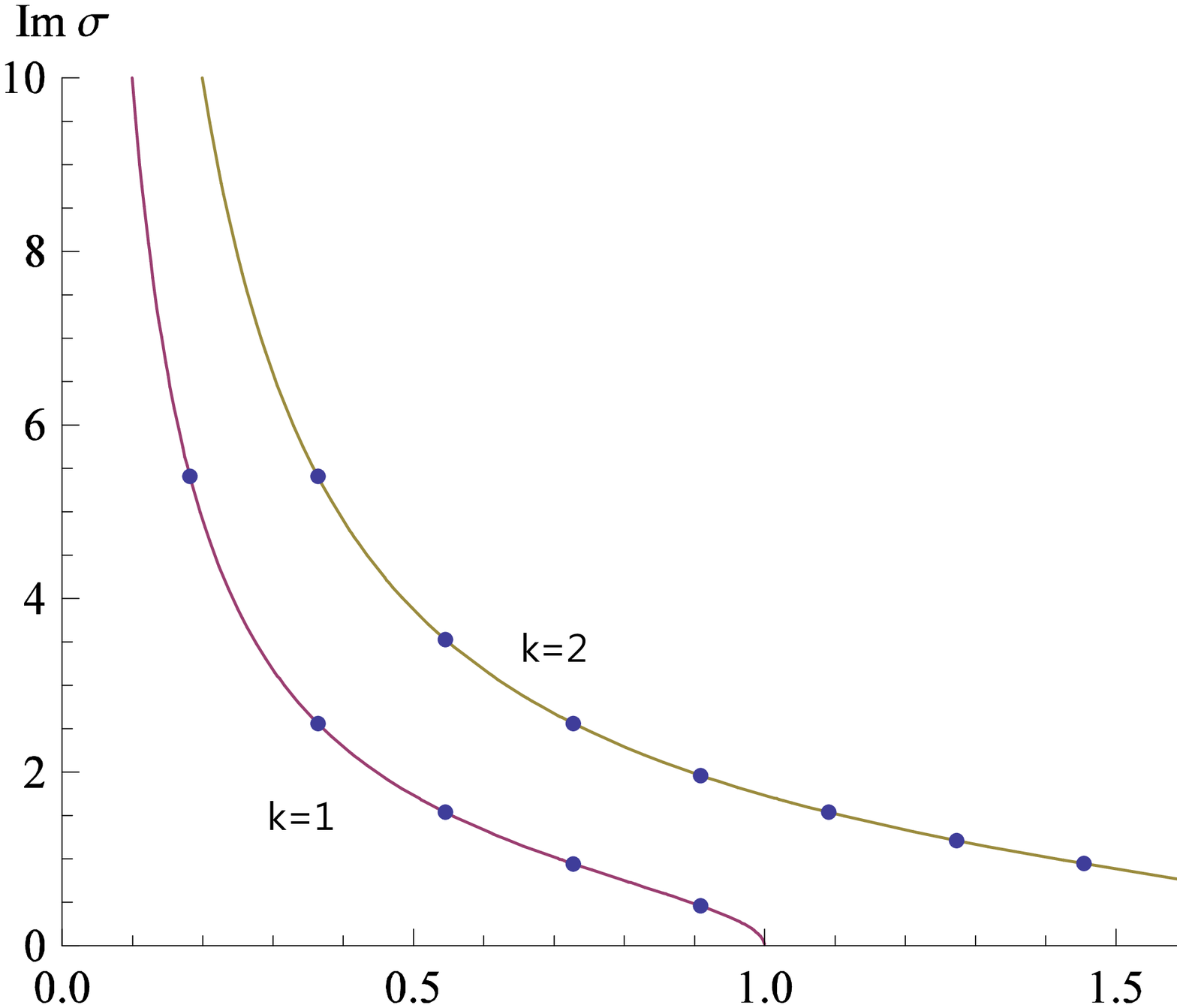}}
\vspace{-2.5cm} \\
\caption{\small The real and imaginary conductivity at $t_0 = 1$ with $a_1 = b_1 \le 1$.}
\label{fcawoc}
\end{center}
\end{figure}

For $a_x$, the Green function becomes
\be \la{defgr}
G_{xx} \equiv \frac{\pa^2 S_B}{\pa a_0 \pa a_0} = i w  \sqrt{1- \frac{k^2 t_0^2}{w^2}} ,
\ee
and the conductivity of the dual system is given by
\be \la{guide}
\s = \frac{G_{xx}}{i w} =  \sqrt{1- \frac{t_0^2 k^2 }{w^2}} .
\ee
For the time-like case ($ w^2 > k^2 t_0^2$), the conductivity is real. In the space-like case,
the imaginary conductivity appears.
In addition, the AC conductivity for $k=0$ becomes a constant $\s_{AC}  = 1$,
in which there is no imaginary part of the conductivity.
In Figure \ref{fcawoc}, we plot the real and imaginary conductivity, in which we can see that
as the momentum $k$ increases the real or imaginary conductivity decreases or increases
respectively. Furthermore, as shown in \eq{guide} and figure 2,
the real and imaginary conductivities become zero at $w^2 = t_0^2 k^2$. Below or above
this critical point, there exists only the imaginary or real conductivity respectively.

\vspace{1cm}
\noindent{\bf ii) $ \half < b_1 < a_1 \le 1 $} \\

It is impossible to solve \eq{maxwell} analytically with arbitrary parameters $a_1$ and $b_1$.
So, instead of solving \eq{maxwell} analytically, we will try to find a Green function
and electric conductivity numerically. To do so, we should first know the perturbative
behavior of $a_x$ near the horizon as well as the asymptotic boundary.

\begin{figure}
\begin{center}
\vspace{1cm}
\hspace{-0.5cm}
\subfigure{ \includegraphics[angle=0,width=0.5\textwidth]{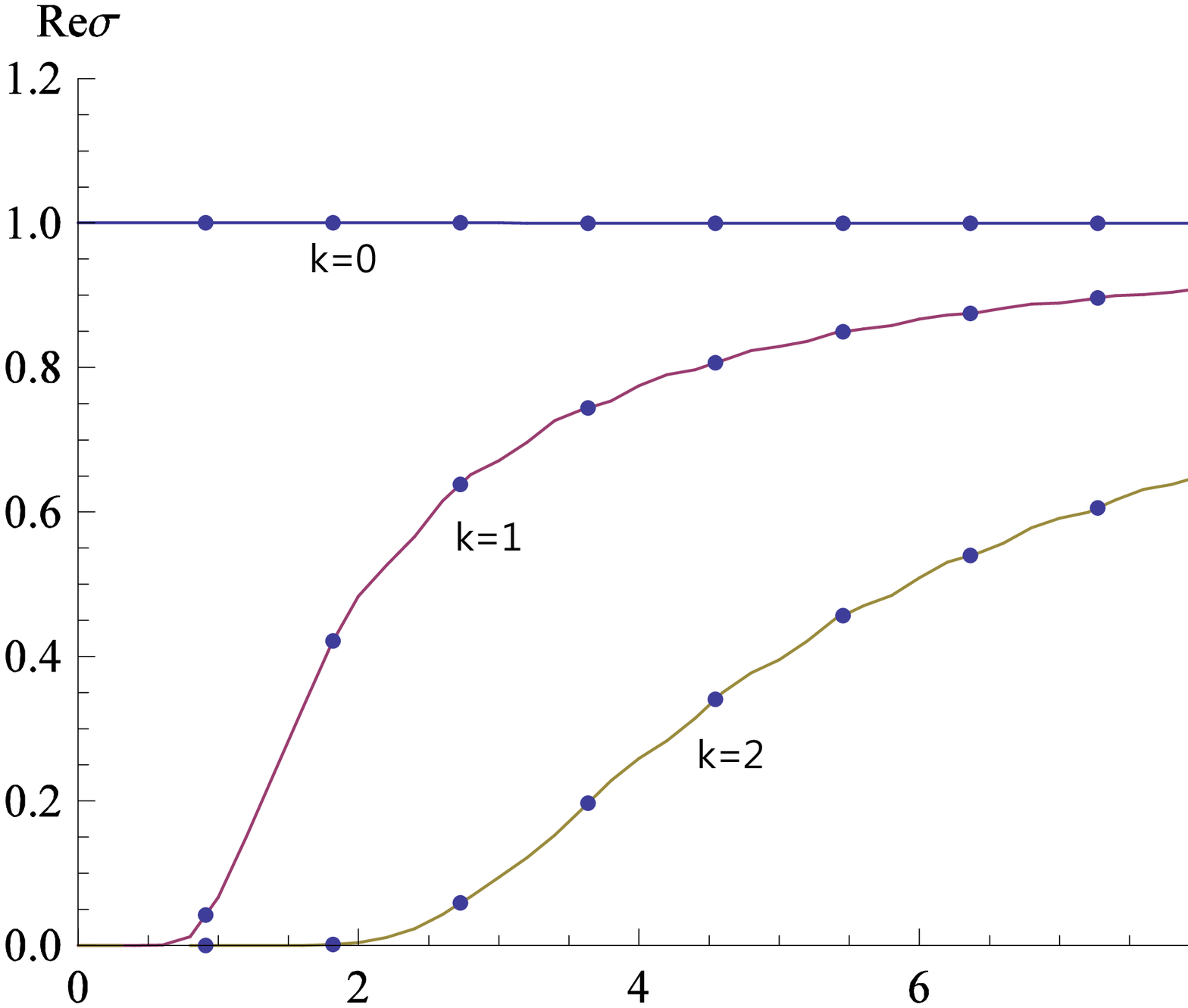}}
\hspace{-0.5cm}
\subfigure{ \includegraphics[angle=0,width=0.5\textwidth]{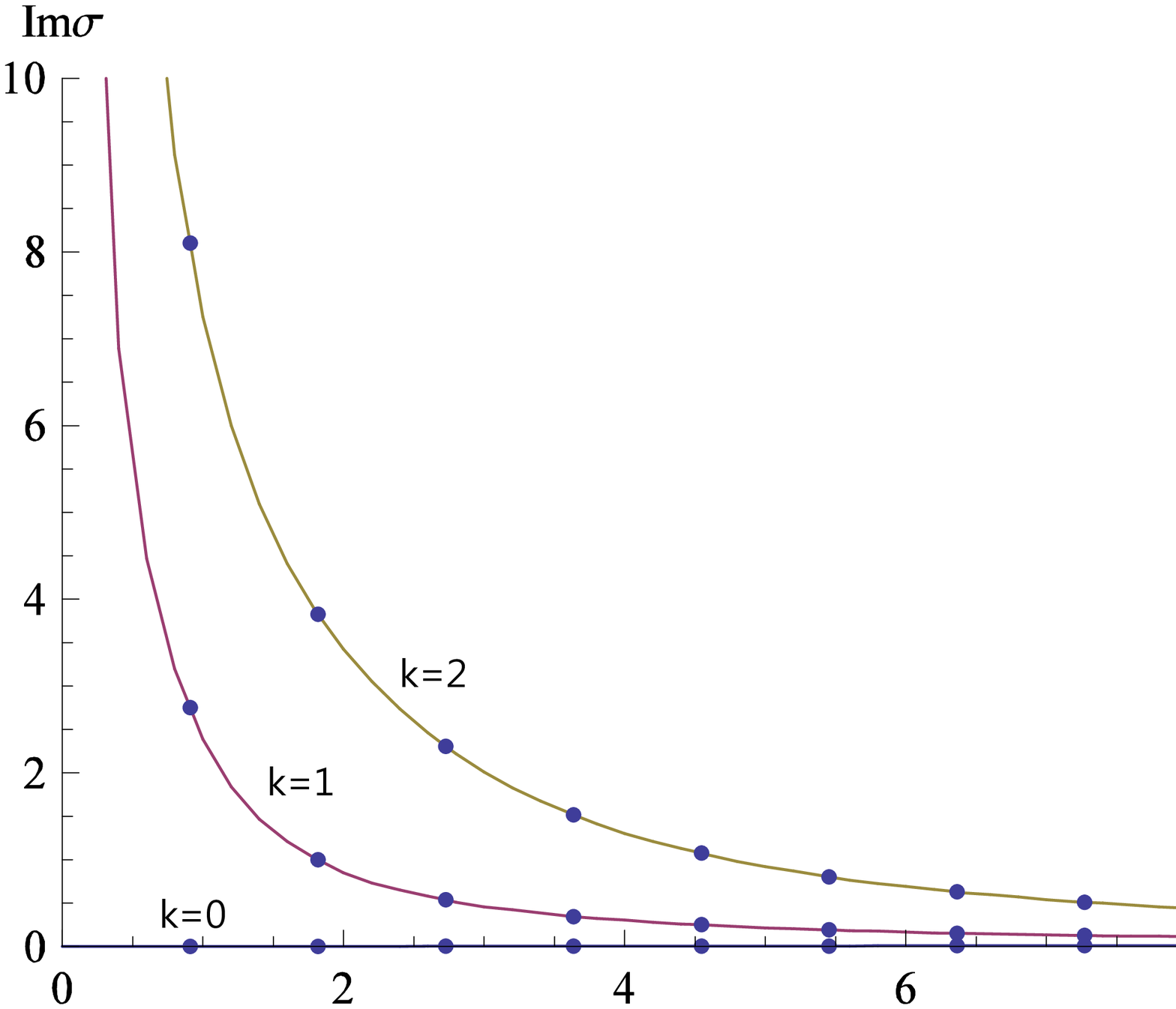}}
\vspace{-2.5cm} \\
\caption{\small The real and imaginary conductivity at $t_0 = 1$ with $\half <b_1  < a_1 \le 1$.}
\label{figg3}
\end{center}
\end{figure}

At the horizon, since $1/z^{4 (1- a_1)}$ term in \eq{maxwell} is dominant,
the approximate solution satisfying the incoming boundary condition is given by
\be \la{hs1}
a_x = c   \exp \ls i \frac{w }{t_0^2 (2 a_1 -1)} z^{ 2 a_1 -1 } \rs .
\ee
Near the boundary, $1/z^{4-2 a_1-2b_1}$ term in \eq{maxwell} is dominant, so \eq{maxwell}
is reduced to
\be \la{eqforax}
0 =  a_x'' + \frac{2(1-a_1)}{z}  a_x' - \frac{k^2}{t_0^2 z^{4-2 a_1-2b_1}} a_x .
\ee
The leading two terms of asymptotic solution are
\be \la{bdsol}
a_x = c_1 + c_2   z^{2 a_1 -1} ,
\ee
where $c_1$ and $c_2$ are integration constants.
To find a relation between
two integration constants and $c$, we should solve \eq{maxwell} numerically
with the initial conditions determined from \eq{hs1}. To control the limit $z \to \infty$
we introduce an IR cut-off $z_0$, which is a very large number. At this IR cut-off, $a_x$ and $a_x'$
become
\bea
a_x (z_0) &=&  c   \exp \ls i \frac{w }{t_0^2 (2 a_1 -1)} z_0^{ 2 a_1 -1 } \rs , \nn
a_x' (z_0) &=&  \frac{i c w \ z_0^{ 2 a_1 -2 }}{t_0^2}
\exp \ls i \frac{w }{t_0^2 (2 a_1 -1)} z_0^{ 2 a_1 -1 } \rs .
\eea
Using these initial values, we can solve \eq{maxwell} numerically and find numerical
values for $a_x (\ep)$ and $a_x' (\ep)$, where $\ep$($\ep \to 0$) means an UV cut-off.
Then, two integration constants $c_1$ and $c_2$ can be determined by
$a_x (\ep)$ and $a_x' (\ep)$
\bea
c_1 &=& a_x (\ep) , \nn
c_2 &=& \frac{a_x' (\ep)}{(2 a_1 -1) \ep^{2 a_1 -2}}.
\eea
If we identify the boundary value of $a_x$ with a source term $a_0$
\be
a_0 = \lim_{\ep \to 0} a_x ( \ep ) .
\ee
$c_1$ and $c_2$ correspond to a source and the expectation value of the boundary
dual operator respectively. Then, from the boundary action the
electric conductivity becomes
\be
\s =  \frac{t_0^2 a_x' (\ep)}{ i w a_x (\ep) \ep^{2 a_1 - 2 }} .
\ee
In Figure \ref{figg3}, we plot the real and imaginary conductivity. Notice that
in this case there is no critical point like the $ \half < a_1 = b_1 \le 1$ case.
In other words, the real and imaginary conductivity is well defined on the whole
range of the frequency. Especially, for large $k$ the real conductivity becomes
zero as the frequency goes to zero. For $k=0$, the real conductivity is a constant
like the previous case. For $k \ne 0$ the conductivity grows as the frequency
increases, which is opposite to the strange metallic conductivity.

\vspace{1cm}
\noindent{\bf iii) $\half <  a_1 < b_1 \le 1$} \\

In this case, $k^2$ term in \eq{maxwell} is dominant at the horizon. Due to the
sign of it, the near horizon behavior of this solution is space-like. So we should
impose the regularity condition instead of the incoming boundary condition. More precisely,
in this parameter region the equation governing $a_x$ at the horizon reduces to
\eq{eqforax}. The exact solution of it is given by
\be
a_x =  z^{a_1 - \half} \ls d_1' I_{- \n} \ls x \rs  + d_2' I_{\n} \ls x \rs  \rs,
\ee
with two integration constants $d_1'$ and $d_2'$, where
$I_{\n} (x)$ is a modified Bessel function and
\be
\n = \frac{2 a_1 -1}{2 (a_1 + b_1 -1)} \qquad {\rm and} \qquad
x = \frac{k z^{a_1 + b_1 -1}}{t_0 (a_1 + b_1 -1)} .
\ee
At the horizon, the leading terms of it become
\be
a_x = \frac{1}{z^{(b_1 - a_1)/2}} \lb d_1 \exp \ls - \frac{k z^{a_1 + b_1-1}}{t_0 (a_1 + b_1 -1) } \rs
+ d_2 \exp \ls \frac{k z^{a_1 + b_1-1}}{t_0 (a_1 + b_1 -1) } \rs \rb ,
\ee
where $d_1$ and $d_2$ are different constants with $d_1'$ and $d_2'$. In the above,
the second term diverges at the horizon $z \to \infty$, so we can pick up the regular solution
by imposing the regularity at the horizon, which is the same as imposing the incoming
boundary condition when $k \to - i k$. From the horizon solution
\be \la{solhor}
a_x = \frac{d_1  \exp \ls - \frac{k z^{a_1 + b_1-1}}{t_0 (a_1 + b_1 -1) } \rs}{z^{(b_1 - a_1)/2}}  ,
\ee
we can determine initial values for $a_x (z_0)$ and $a_x' (z_0)$
at the IR cut-off $z_0$.
The near boundary solution in $\half <  a_1 < b_1 \le 1$ is also given by the same form
in \eq{bdsol}.

\begin{figure}
\begin{center}
\vspace{1cm}
\subfigure{ \includegraphics[angle=0,width=0.5\textwidth]{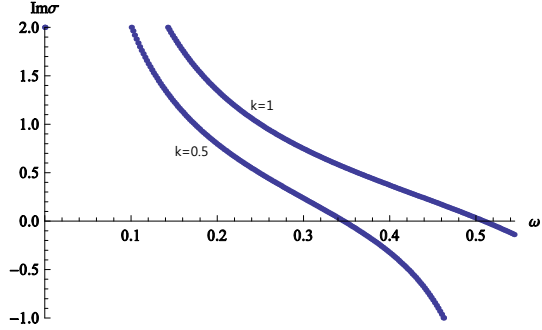}}
\vspace{-2.5cm} \\
\caption{\small The imaginary conductivity at $t_0 = 1$ with $\half < a_1  < b_1 \le 1$.}
\label{fig4}
\end{center}
\end{figure}

Before calculating the Green function and electric conductivity, notice that
$a_x$ in \eq{solhor} is real due to the regularity condition at the horizon.
So the resulting Green function is also real, which implies that the
conductivity is a pure imaginary number.
Following the technique explained in the previous section, the
unknown integration constants $c_1$ and $c_2$ can be determined by the boundary values
$a_x (\ep)$ and $a_x'(\ep)$ after solving \eq{maxwell} numerically. Figure \ref{fig4}
shows the imaginary conductivity.

\subsubsection{for $a_1 >1$}

\begin{figure}
\begin{center}
\vspace{1cm}
\hspace{-0.5cm}
\subfigure{ \includegraphics[angle=0,width=0.5\textwidth]{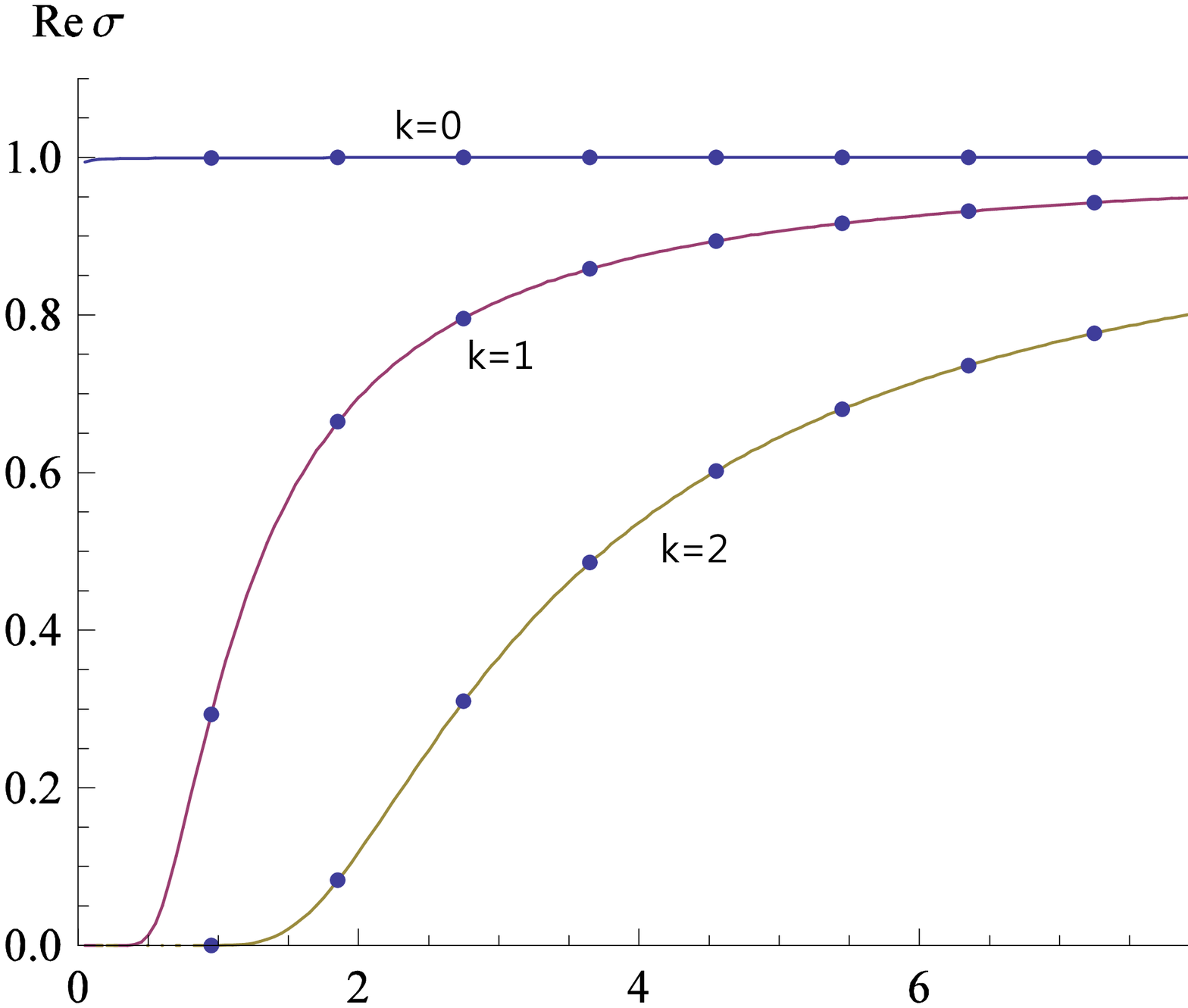}}
\hspace{-0.5cm}
\subfigure{ \includegraphics[angle=0,width=0.5\textwidth]{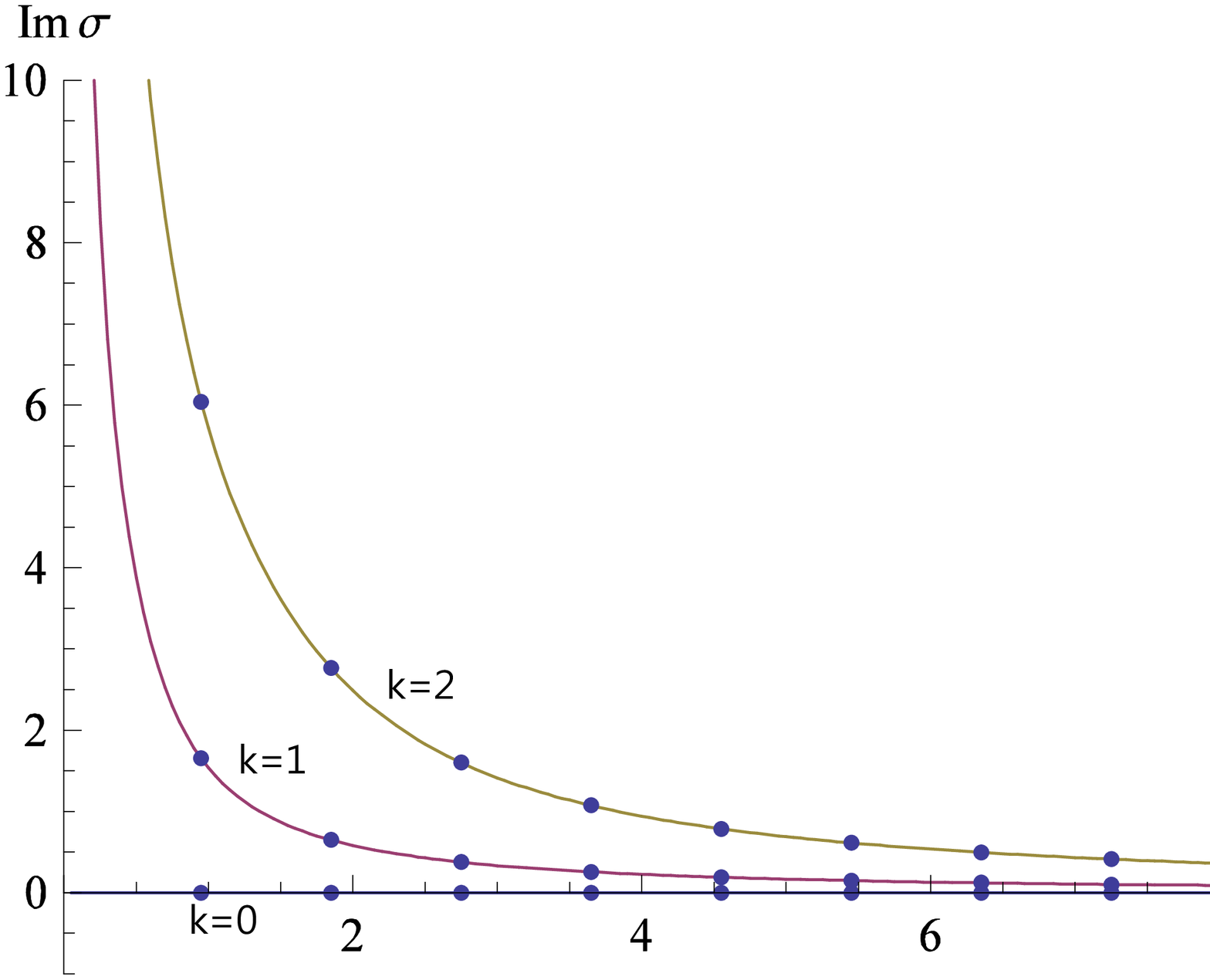}}
\vspace{-2.5cm} \\
\caption{\small The real and imaginary conductivity when $a_1 = 5/4$, $b_1 = 3/4$ and $t_0 = 1$ .}
\label{zeroa1beb1}
\end{center}
\end{figure}

In this case, the equation for $a_x$ becomes
\be \la{diffeq}
0 = a_x'' - \frac{2 (a_1 - 1)}{z}  a_x' + \ls \frac{w^2}{t_0^4} z^{4(a_1-1)}
-\frac{k^2}{t_0^2 } z^{2 (a_1 + b_1 -2)} \rs a_x .
\ee
At the horizon($z \to \infty$), the first term proportional to $w^2$ is dominant, so the
approximate solution satisfying the incoming boundary condition is given by
\be \la{hosol}
a_x \approx  \ \exp \ls \frac{i w z^{2 a_1 -1}  }{t_0^2 (2 a_1 - 1)} \rs .
\ee
For $k=0$, the above is an exact solution satisfying the incoming boundary
condition, which is the same as one in \eq{soexs} with $k=0$. In this case, from
the result in \eq{guide} we can easily find that the conductivity becomes $1$.
Notice that in Figure \ref{zeroa1beb1} the numerical calculation for the conductivity at $k=0$ gives
the same result obtained by analytic calculation.

Near the boundary ($z \to 0$), the $k^2$ term is dominant,
so the approximate solution becomes
\be \la{bdsol0}
a_x \approx a_0 \ls 1 + c \ z^{2 a_1 -1} \rs  ,
\ee
where $a_0$ is the boundary value of $a_x$. To determine a constant $c$,
we numerically solve the differential equation in \eq{diffeq} with the initial values
given at the horizon. Then, we can find the values for $a_x$ and $a_x'$ at the boundary.
Comparing these results with \eq{bdsol0}, we can determine the unknown constant $c$
numerically. Using these results, the Green functions
and conductivity are given by
\bea
G_{xx} &=& t_0^2 (2 a_1 -1) c , \nn
\s &=& \frac{G_{xx}}{i w} .
\eea
In Figure \ref{zeroa1beb1}, we present several conductivity plots depending on the
momentum, which is very similar to one obtained in the $\half < b_1 < a_1 \le 1 $
case. Notice that for $\half < b_1 < a_1 \le 1 $ and $a_1 >1$ the DC conductivity
is zero for $k = 1$ or $2$.

\subsection{At finite temperature}

First, we consider the simplest case $k=0$ in which we can obtain an exact solution.
To solve \eq{maxwell}, we introduce a new coordinate
\be
d u = \frac{dz}{z^{2 - 2 a_1} f} .
\ee
Then, in the $u$-coordinate \eq{maxwell} for $k=0$ reduces to
\be \la{eqk0ft}
0 = \pa_u^2 a_x + \frac{w^2}{t_0^4} a_x ,
\ee
where the relation between $u$ and $z$ is given by
\be
u = \frac{z^{2 a_1 -1}}{2 a_1 -1} \  _2 F_1 \ls \frac{2 a_1  -1}{2 a_1 + 2 b_1 -1} ,
\frac{2 (2 a_1 + b_1 -1)}{2 a_1 + 2 b_1 -1}, z^{2 a_1 + 2 b_1 -1} \rs .
\ee
Near the horizon ($z \sim z_h$), the above relation
becomes
\be
u \approx - \frac{z_h^{2 a_1 - 1}}{2 a_1 + 2 b_1 -1} \log (z_h - z) ,
\ee
so the horizon lies at $u = \infty$ for $a_1 > 1/2$. Near the boundary
($z \to 0$), $u$ is related to $z$
\be
u \approx \frac{z^{2 a_1 -1}}{2 a_1 - 1} .
\ee
At the horizon, the solution of \eq{eqk0ft} satisfying the incoming boundary condition
is given by
\be
a_x = a_0 \exp \ls i \frac{w u}{t_0^2}  \rs ,
\ee
where $a_0$ is the boundary value of $a_x$. Near the boundary, the expansion of the
above solution becomes
\be
a_x =a_0 \ls 1 + i \frac{w}{(2 a_1 -1) t_0^2} z^{2 a_1 -1}  \rs ,
\ee
which is the same as one in the zero temperature case. Therefore, the Green function
and conductivity for $k=0$ is the same as the zero temperature result.

Now, we consider the general case with non-zero $k$. In this case, it is very difficult
to find an analytic solution, so we adopt a numerical method.
Since the black hole factor $f$ becomes zero at the horizon ($z \to z_h$),
the differential equation in \eq{maxwell} near the horizon reduces to
\be
0 =  a_x'' +\frac{f'}{f} a_x' + \frac{w^2}{t_0^4 z^{4(1- a_1)} f^2} a_x ,
\ee
where
\be
f = 1 - \frac{z^{2 a_1 + 2 b_1 -1}} {z_h^{2 a_1 + 2 b_1 -1}} .
\ee
The leading term of the solution satisfying the incoming boundary condition is
\be
a_x = c f^{- i \n} \ls 1 + \cdots \rs ,
\ee
with
\be
\n = \frac{w \ z_h^{2 a_1 -1 }}{{(2 a_1 + 2 b_1 -1) \ t_0^2}} ,
\ee
where the ellipsis implies higher order terms and $c$ is an integration
constant. Since at $z=z_h$ $a_x$ and $a_x'$
become zero, we pick up the initial values at $z_i = z_h - \ep$, where $\ep$ is
a very small number, for solving the differential equation numerically.
After solving \eq{maxwell}, we can find numerical values
for $a_x (\ep)$ and $a_x' (\ep)$, where $\ep$ is a small number,
up to multiplication constant $c$ at the boundary.

\begin{figure}
\begin{center}
\vspace{1cm}
\hspace{-0.5cm}
\subfigure{ \includegraphics[angle=0,width=0.5\textwidth]{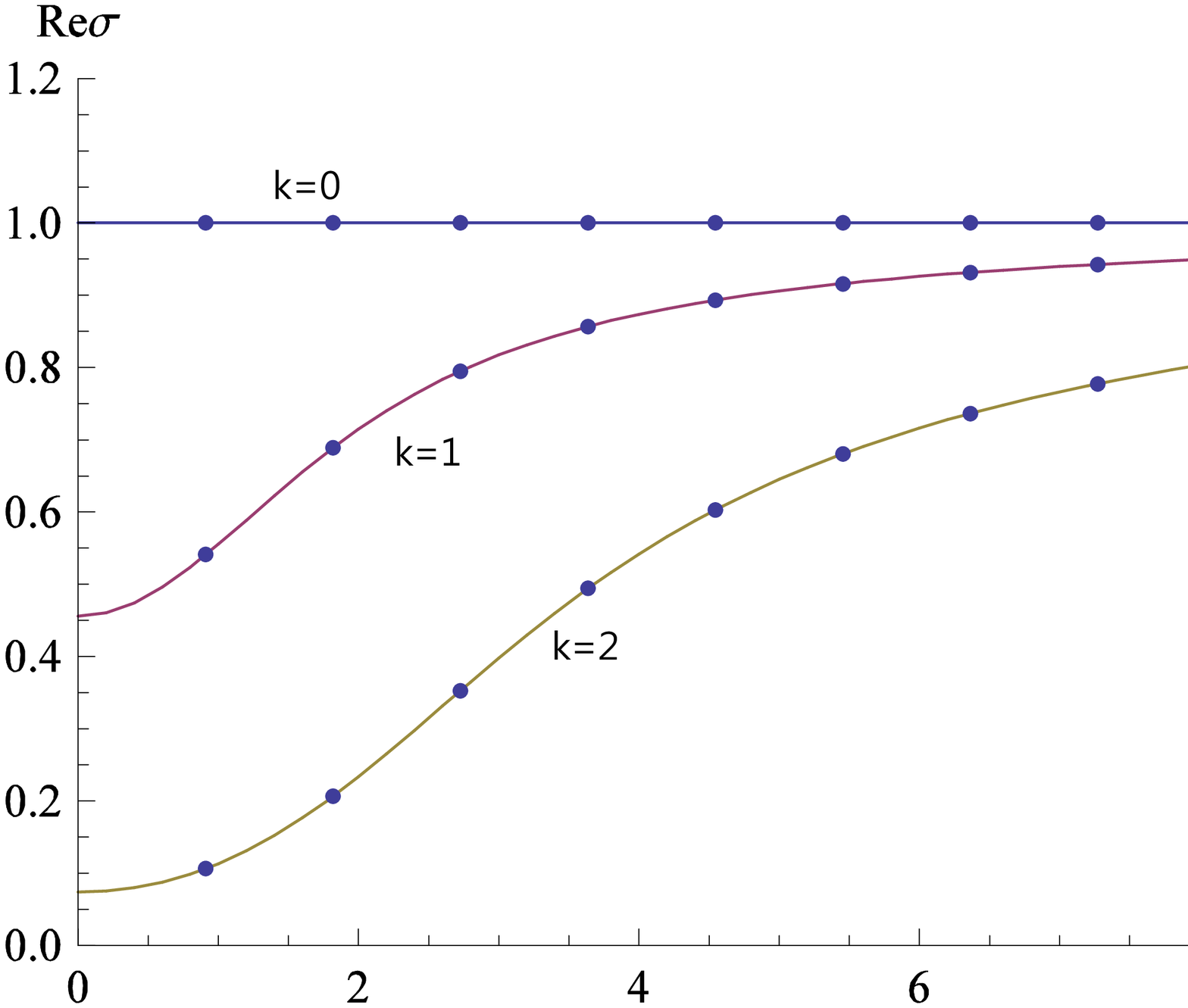}}
\hspace{-0.5cm}
\subfigure{ \includegraphics[angle=0,width=0.5\textwidth]{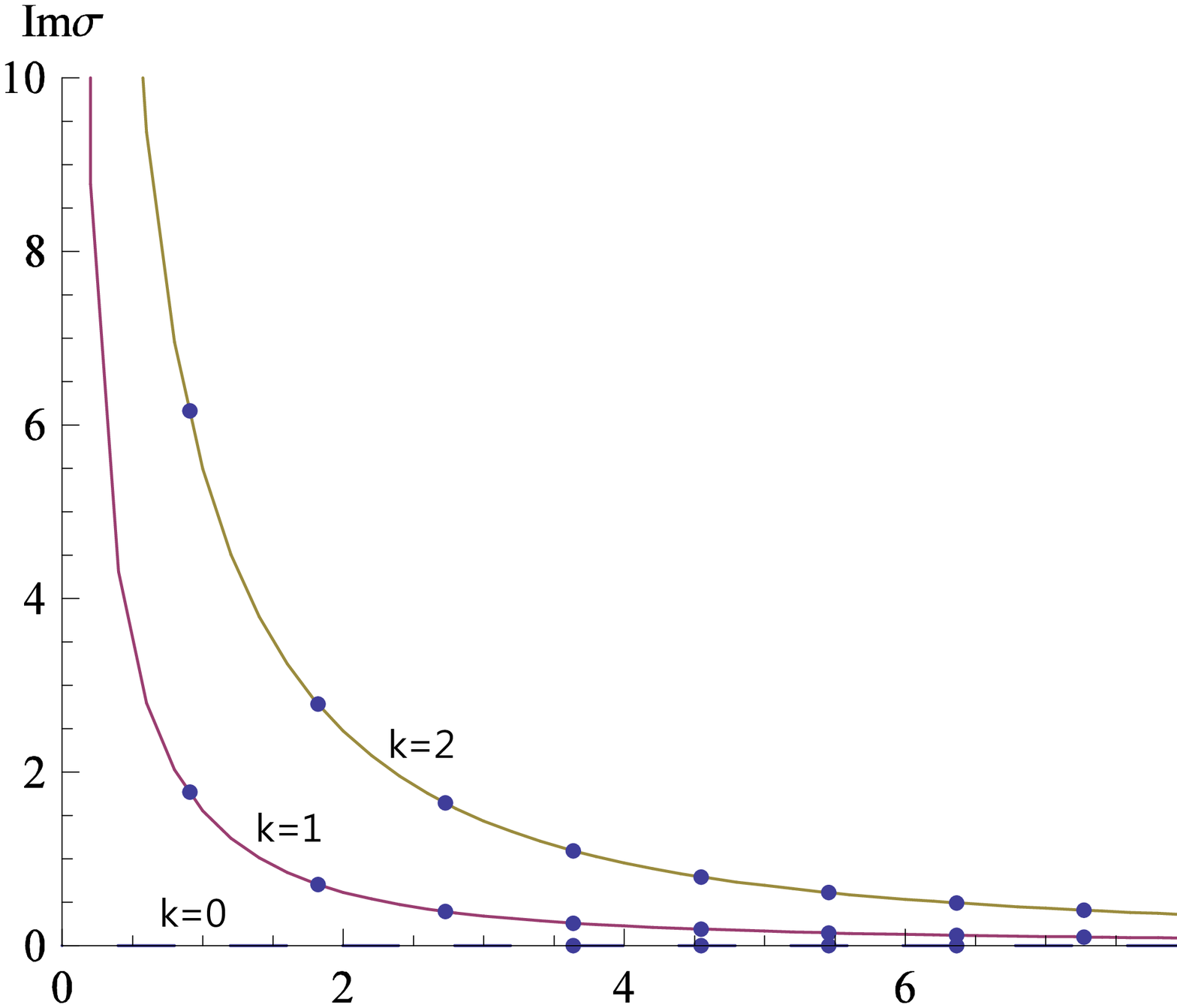}}
\vspace{-2.5cm} \\
\caption{\small The real and imaginary conductivity at finite temperature $T = 0.239$ ($z_h=1$)
when $a_1 = 5/4$, $b_1 = 3/4$ and $t_0 = 1$ .}
\label{fig6}
\end{center}
\end{figure}

\begin{figure}
\begin{center}
\vspace{1cm}
\hspace{-0.5cm}
\subfigure{ \includegraphics[angle=0,width=0.5\textwidth]{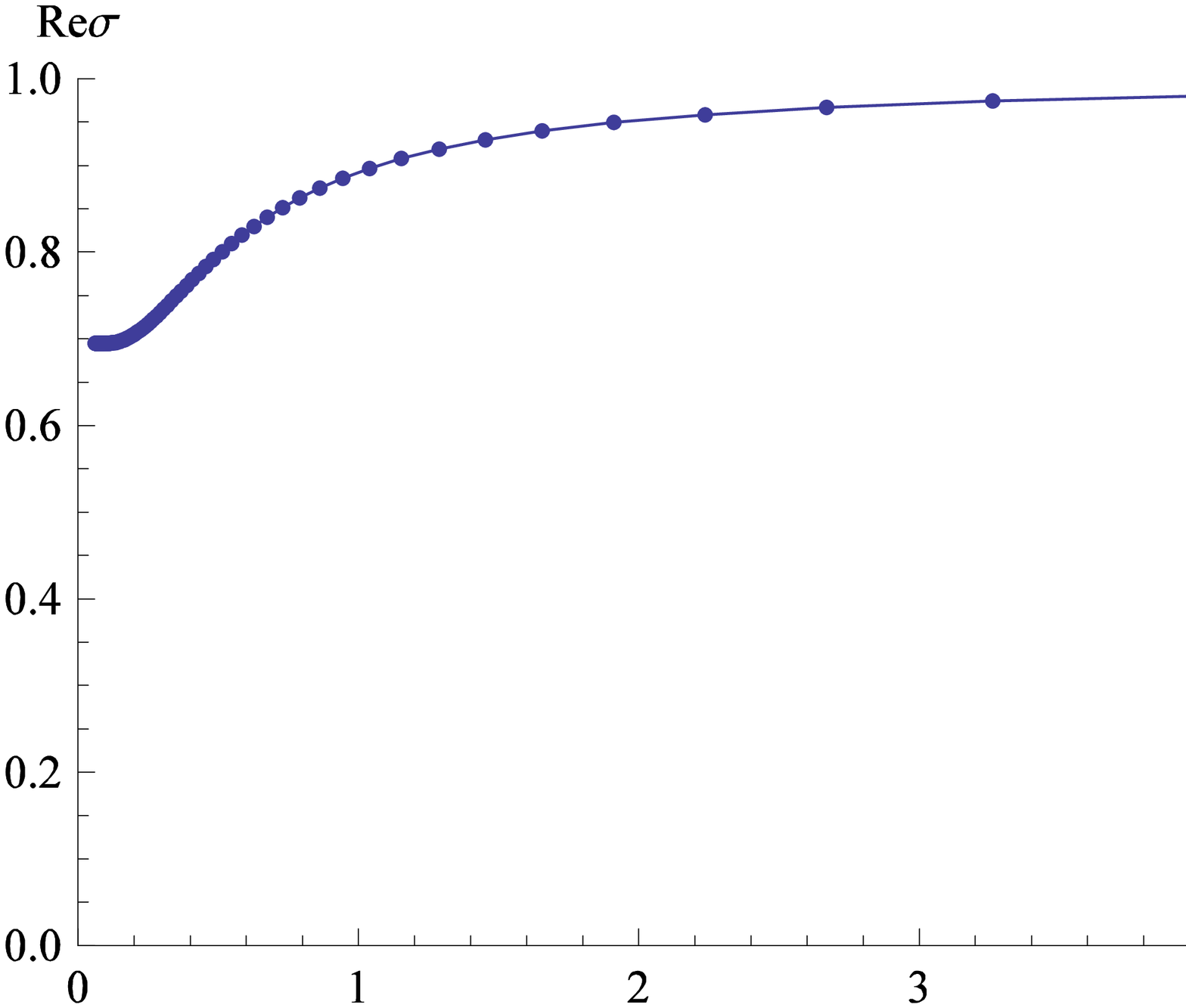}}
\hspace{-0.5cm}
\subfigure{ \includegraphics[angle=0,width=0.5\textwidth]{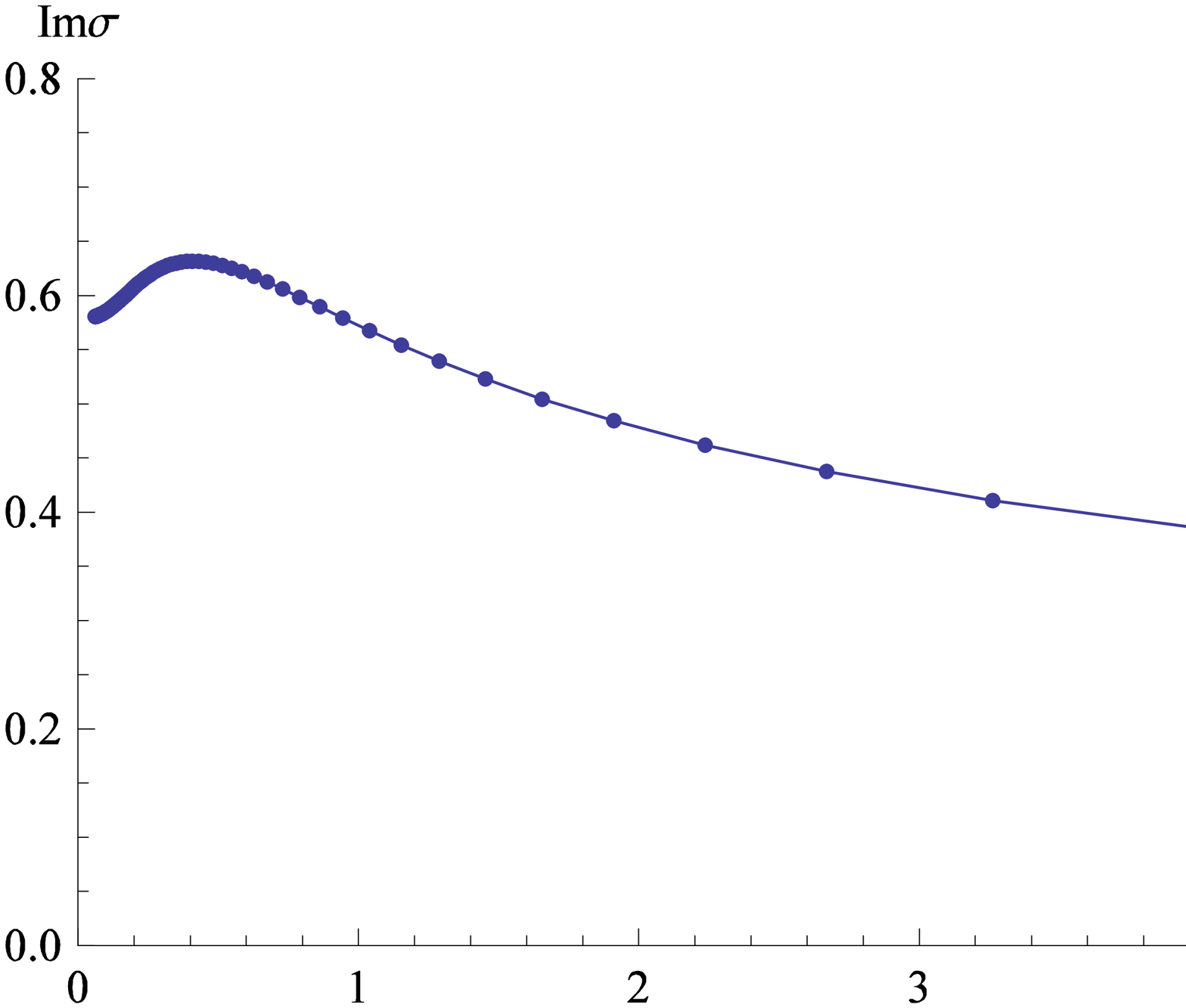}}
\vspace{-2.5cm} \\
\caption{\small The real and imaginary conductivity depending on temperature when
$w=2$, $k=1$, $a_1 = 5/4$, $b_1 = 3/4$ and $t_0 = 1$ .}
\label{fig7}
\end{center}
\end{figure}

To understand the boundary behavior of $a_x$, we should know the perturbative form
of $a_x$ near the boundary. From \eq{maxwell}, the perturbative solution near the boundary
is given by
\be \la{bdsol1}
a_x = a_0 \ls 1 + A \ z^{2 a_1 -1} \rs ,
\ee
where $a_0$ is the boundary value of $a_x$ and $A$ is an integration constant, which
will be determined later.
Using the above numerical results $a_x (\ep)$ and $a_x' (\ep)$,  $a_x$ can be written as
\be
a_x (z) = c \ls a_x (\ep) + \frac{a_x' (\ep)}{2 a_1 -1} \
\frac{z^{2 a_1 -1}}{\ep^{2 a_1 -2}} \rs ,
\ee
Matching this value with \eq{bdsol1} gives $ c = \frac{a_0}{a_x (\ep)} $ and $A$ becomes
\be
A = \frac{a_x' (\ep)}{ (2 a_1 -1) \ep^{2 a_1 -2} a_x (\ep)} .
\ee
Therefore, the Green function and conductivity can be written as
\bea
G_{xx} &=& (2 a_1 - 1) t_0^2 A  , \nn
\s &\equiv& \frac{G_{xx}}{i w} =\frac{ t_0^2 }{ i w \ep^{2 a_1 -2}}
\frac{a_x' (\ep)}{ a_x (\ep)}  .
\eea
Figure \ref{fig6} shows the conductivity at finite temperature.
Similarly to the zero temperature cases, for $k=0$ the real conductivity is still
a constant. But for $k=1$ or $2$, the finite temperature conductivity goes to a constant
as the frequency goes to zero,
while the zero temperature one approaches to zero.
This implies that the finite temperature DC conductivity is a constant while
the zero temperature DC conductivity is zero. We can also see that like the zero
temperature case the conductivity grows as the frequency increases,
which is different with the strange metallic behavior.
Moreover, since our background geometry
is not a maximally symmetric space, the dual boundary theory is not conformal. So we can
expect that there exists the non-trivial temperature dependence of the conductivity.
In Figure \ref{fig7} we plot the electric conductivity depending on temperature.

\section{Discussion}

There exist two complementary approaches for studying
applications of the gauge/gravity duality to condensed matter systems. One is
the bottom-up approach in which on a given gravity solution we investigate
the dual field theory. The other is the
top-down approach in which, after considering the probe brane configuration on the geometry
obtained from the string theory, the dual field theory is investigated by studying various fluctuations
on the probe brane.  In
this paper we mainly adopt the bottom-up approach. For the top-down approach,
see Ref. \cite{lpp}.

Due to the peculiar properties, charged dilaton black holes may
provide new backgrounds for describing the gravity duals of certain
condensed matter systems. In this paper we studied conductivities of
charged dilaton black holes with a Liouville potential both at
zero and finite temperature. In general, the anisotropic (charged dilaton) black hole
has three free parameters. Depending on which parameters we choose, the
anisotropic background can reduce to $AdS_4$, $AdS_2 \times S^2$ and the Lifshitz-like
space. To investigate the dual theory which may describe
some aspects of the condensed matter system, we have calculated its electric conductivity in
various parameter regions.

First, we have considered the gauge fluctuation of the background gauge field,
which is coupled with the dilaton field. Due to this non-trivial dilaton coupling,
the conductivity of this system depends on the frequency non-trivially.
After choosing appropriate parameters, at both zero and finite temperature
we obtained the strange metal-like AC conductivity
proportional to the frequency with a negative exponent.
Because our model have
three free parameters, it may shed light on investigating other dual field theory
having more general exponent.

Second, we have also investigated another dual field theory whose
dual gravity theory has the another U(1) gauge field fluctuation without
the dilaton coupling. Due to the parameters in the original action, there are several
parameter regions giving different conductivities. We have classified
all possible conductivities either analytically or numerically. Here, we found that
the conductivities for $k=0$ at zero and finite temperature become a constant
because there is no dilaton coupling with the new gauge field fluctuation.
This implies that to describe the non-trivial conductivity depending on the frequency
like the strange metal, it would be important to consider
the dilaton coupling effect at least in the bottom-up approach.
We have also investigated the conductivity depending on the spatial
momentum and the temperature. As the spatial momentum increases we found that the
real conductivity goes down. In addition, we found that the finite temperature
DC conductivity becomes a non-zero constant while the zero temperature one is zero in
this set-up.

One further generalization is to incorporate the magnetic field and
to find dyonic black hole solutions. Once we find such exact
solutions carrying both electric and magnetic charges, it would be
interesting to study the thermodynamics and transport coefficients,
such as the Hall conductivity~\cite{Hartnoll:2007ai}, in the
presence of the magnetic field. Another interesting direction is to
study the non-Fermi liquid behavior in the solutions we obtained,
following~\cite{Lee:2008xf, Liu:2009dm, Cubrovic:2009ye,
Faulkner:2009wj}. We hope to report some results about
such fascinating topics in the future. \\

\vspace{1cm}

\bigskip \goodbreak \centerline{\bf Acknowledgements}

This work was supported by the National Research Foundation of Korea
(NRF) grant funded by the Korea government (MEST) through the Center
for Quantum Spacetime (CQUeST) of Sogang University with grant
number 2005-0049409. C. Park was also
supported by Basic Science Research Program through the
National Research Foundation of Korea(NRF) funded by the Ministry of
Education, Science and Technology(2010-0022369).

\vspace{1cm}



\end{document}